\documentclass[prd,amsmath,amssymb,longbibliography,superscriptaddress,twocolumn,nofootinbib,10pt]{revtex4}

\usepackage{dcolumn}
\usepackage{bm}
\usepackage{amssymb}

\usepackage{booktabs}
\usepackage{amsmath}
\usepackage{multirow}
\usepackage{graphicx} 
\RequirePackage{xcolor}

\newcommand{\cfootnote}[2][red]{%
    {\color{#1}\footnote{#2}}%
}
\usepackage[colorlinks,linkcolor=black,anchorcolor=black,citecolor=blue,hyperfootnotes=true]{hyperref}
\newcommand{\beq}[1]{\begin{equation}\label{#1}}
 \newcommand{\eeq}{\end{equation}}
 \newcommand{\bea}{\begin{eqnarray}}
 \newcommand{\eea}{\end{eqnarray}}

\newcommand{\Mpc}{\mathrm{~km~s^{-1}~Mpc^{-1}}}

\newcommand{\gp}{\gamma_{\rm{PPN}}}
\def\({\left(}
\def\){\right)}
\def\[{\left[}
\def\]{\right]}

\def\aj{Astron. J.}
\def\aap{Astron. Astrophys.}
\def\mnras{Mon. Not. Roy. Astron. Soc.}
\def\apjl{Astrophys. J. Lett.}
\def\jcap{JCAP}
\def\nar{New Astronomy Reviews}
\def\apjs{Astrophys. J., Suppl. Ser.}

\begin{document}

\title{A robust test of general relativity at the galactic scales by combining strong lensing systems and gravitational wave standard sirens }

\author{Tonghua Liu}
\affiliation{School of Physics and Optoelectronic Engineering, Yangtze University, Jingzhou, 434023, China;}
\author{Marek Biesiada}
\affiliation{National Centre for Nuclear Research, Pasteura 7, PL-02-093 Warsaw, Poland;}
\author{Shuxun Tian}
\affiliation{Department of Astronomy, Beijing Normal University, 100875, Beijing, China;}

\author{Kai Liao}
\email{liaokai@whu.edu.cn;}
\affiliation{School of Physics and Technology, Wuhan University, Wuhan 430072, China;}

\begin{abstract}
The measurement of the parametrized post-Newtonian parameter $\gp$ is a robust test of general relativity (GR). In some modified theories of gravity, $\gp$  may evolve with the redshift and deviate from one at high redshifts. This means that precise constraints on $\gp$ acquired in the solar system experiments could not be sufficient to test such theories and it is necessary to constrain $\gp$ with high precision at high redshifts.  However, in many approaches aimed at extragalactic tests of GR, the results might be biased due to entanglement of various factors, such as cosmic curvature, cosmic opacity,  and the Hubble constant. Strong lensing systems naturally provide a laboratory to test  $\gp$ at galactic scales and high redshifts, but there is degeneracy between measured strength of gravity and cosmic distances in the lensing system. Gravitational waves (GWs) from binary neutron star mergers (standard sirens) provide a direct way to break this degeneracy by providing self-calibrated measurements of the luminosity distance.  {We investigate the possibility of estimating $\gp$ by combining well measured strongly lensed systems with GW signals from coalescing neutron stars. Such  combination provides a cosmological-model independent, relatively pure and unbiased method for the inference of $\gp$ parameter, avoiding the influence of the above factors and the mass-sheet degeneracy in the lens.} Based on the simulated future 55 lensed quasar systems we demonstrated that the precision of $\gp$ parameter obtained by our method could be of order of $\sim 10^{-2}$. One may reasonably expect that our approach will play an increasingly important role in precise testing the validity of general relativity at galactic scales and high redshifts.
\end{abstract}

\maketitle

\section{Introduction}           

\label{sect:intro}
Over the past few decades, observations of Type Ia supernovae (SNe Ia) have revealed that the expansion of the Universe is accelerating \cite{1998AJ....116.1009R,1999ApJ...517..565P}. Based on Einstein's theory of General Relativity (GR) \cite{1984ucp..book.....W}, and assuming that the Universe is homogeneous and isotropic on large scales \citep{1972gcpa.book.....W}, it is generally accepted that
the so called $\Lambda$CDM model correctly describes our Universe.
%
%
Although this model is currently taken as the concordance cosmological scenario and supported by the vast majority of astronomical observations \citep{2016A&A...594A..13P}, it still faces some problems. First, there is a significant tension in the Hubble constant $H_0$ between the cosmic microwave background (CMB) measurement obtained within GR$+\Lambda$CDM using the \emph{Planck} data, which  yielded $H_0=67.4 \pm 0.5 \Mpc$ at the $68\%$ confidence level (CL) \citep{2020A&A...641A...6P} and the value of $H_0=73.2 \pm 1.3 \Mpc$ at the $68\%$ CL reported by SH0ES (\textit{Supernova $H_0$ for the Equation of State}) collaboration using SNe Ia calibrated by local Cepheid variable stars \citep{SH0ES}. The $H_0$ Lenses in COSMOGRAIL's Wellspring (H0LiCOW) collaboration recently reported the measurement $H_0=73.3^{+1.7}_{-1.8}$ $\Mpc$ with a joint analysis of six lensed quasars for a flat $\Lambda$CDM model \citep{2020MNRAS.498.1420W} using a technique independent of above mentioned methods.
Further, the same happens with measurements of cosmic curvature \citep{2020NatAs...4..196D,2021PhRvD.103d1301H} and the growth of structure $S_8$ parameter, which also reveal tension (see a recent review \citep{2022JHEAp..34...49A} for more details) between alternative approaches and analyses. These inconsistencies are the fairly challenging problem in astrophysics and cosmology.  From the theoretical point of view, cosmological constant $\Lambda$ responsible for the accelerating expansion of Universe should be invoked in the framework of GR, but inconsistency between its observed value and theoretical predictions from the quantum field theory
is considerable \citep{Weinberg89}.

Considering the fact that, the values of $H_0$ inferred from \emph{Planck} data and H0LiCOW collaboration are based on GR plus $\Lambda$CDM model, it opens a discussion of whether the GR could fail at larger, cosmological scales. Although GR has passed  with a very high precision all tests at the millimeter scale in the laboratory, up to to the solar system scales  \citep{Ashby02,Bertotti03}, and the detection of gravitational waves also provided the possibility for testing the validity of GR on very extreme scales
\citep{2016ApJ...829...55W,2019PhRvL.123l1101I,2021PhyOJ..14..173S,2021ApJ...921L..19D,2013LRR....16....7G}, the long-range nature of gravity on the extragalactic or cosmological scale is still relatively insufficiently tested.
At present, precise constraints regarding the parametrized post-Newtonian parameter $\gp$ can be obtained at the Solar System or stellar scales in the local Universe, e.g., $\Delta\gp\sim 10^{-5}$ given by Cassini mission \citep{Bertotti03}. However, the extragalactic constraints on this parameter are much weaker, e.g., $\gp=0.97\pm0.09$ on kiloparsec scale by using a nearby lens, ESO 325-G004 given by \citep{Collett2018.Science.360.1342}. Recent reviews of the progress in experimental testing of GR can be found in \citep{Will2014,2019LRR....22....1I,2010GReGr..42.2219U}.
The issue of whether general relativity breaks down on larger cosmological scales should be further validated \citep{Koyama2016}.



Independent alternative techniques could provide new perspectives. As one of the most important predictions of GR, strong gravitational lensing, especially 
strong lensing time-delay measurements, have become a powerful tool for studying cosmology and gravity. In addition, lenses with measured time-delay have recently been found to be more powerful  cosmological probes with capability to measure the angular diameter distances to the lenses $D^A_d$ by combining the measurements of time-delays, image configuration and stellar kinematics (spectroscopy of lens galaxy). The possible deviations from GR enter into the lens formula, which further affects the measured $D^A_d$. However, the possible effects of GR deviation and measured distances are highly degenerate. In other words,
the gravitational mass is inferred from the Einstein radius of the lens, but the observed Einstein radius also depends on the distances in the optical system.
Fortunately, gravitational wave (GW) signals from inspiraling and merging compact binaries provide absolute distance (luminosity distances $D^L$)  \citep{Schutz1986} and do not suffer from interstellar extinction effects, which can help to break aforementioned degeneracy. Such combination of time-delay lenses and GWs opens a new possibility to test deviations from GR at galactic scales and high redshifts in the limit of weak gravitational field.
This letter is organized as follows: In Sec. 2, we present the methodology and observational data. The results and discussion are given in Sec. 3. Finally, we summarize our findings in Sec. 4. The natural units of $c=G=1$ are adopted throughout this letter.  The fiducial cosmological model, a flat $\Lambda$CDM with $\Omega_m=0.30$ and  $H_0=70$ $\Mpc$ is assumed for simulating the lensing data and GW data.


\section{Methodology}\label{sec2}
\subsection{Background}

In the weak field limit, background FLRW metric perturbed by the presence of the lens can be written as
\begin{equation}
ds^2 = -\big(1 + 2 \Phi\big) dt^2 + a(t)^2\big(1 - 2 \Psi\big)d \mathbf{r}^2,
\end{equation}
where $a(t)$ is the scale factor,  $\Phi$ is the Newtonian gravitational potential, and  $\Psi$ is the spatial curvature potential.  We remark here that $\Phi$ and $\Psi$ may be different in some modified gravity theories, i.e., $\Phi$ is gravitational potential that responds to the motion of non-relativistic matter such as baryons and dark matter. The relationship of $\Psi$ to the Newtonian potential is determined by the matter content (e.g., usual matter species in the concordance model plus possible extra scalar fields) and gravity itself.  The deviation from GR is quantified by the ratio $\gp = \Psi / \Phi$ (denoted as the PPN parameter), with $\gp =1$ in the framework of GR. In modified gravity theories, $\Phi$ and $\Psi$ could be unequal with each other \citep{Will2014}, and even their ratio could be a function of space \citep{Hu2007.PRD.76.064004,Saaidi2011.PRD.83.104019, Thomas2023.JCAP.04.016} and time
\citep{2019PhRvD..99f4044T}.  In this letter, we aim to construct an unbiased method to constrain $\gp$  at galactic scales and high redshifts. For simplicity, we assume that $\gp$ is independent of the space coordinates, and only depends on time. Such case can be realized in many modified gravity theories, e.g., the nonlocal Gauss-Bonnet gravity.

Strong lensing systems enables to perform two types of mass measurements: gravitational mass of the lens inferred from the lensed images of the source, and dynamical mass obtained from spectroscopic measurements of stellar kinematics of the lensing galaxy. The dynamical mass is sensitive to the Newtonian gravitational potential $\Psi$ only.
On the other hand, gravitational mass is sensitive to both potentials, i.e. to the Weyl potential (defined as $\Phi_{+}=\frac{\Phi+\Psi}{2}=\frac{1+\gp}{2}\Phi$). Thus, strong lensing  provides a natural laboratory to test gravity and further measure the PPN parameter $\gp$ by directly comparing the difference between the dynamical mass and gravitational mass or comparing $\Phi$ and $\Phi_{+}$.

\subsection{Angular diameter distance from strong lensing}

Combination of time delay and stellar kinematics measurements, can provide the measurement of the angular diameter distance $D_d^A$
to the lens (deflector).  Let us briefly outline the standard procedure for determination of $D_d^A$ pursued in the H0LiCOW program.
For a typical strong lensing system, with quasar at the redshift $z_s$ acting as a background source, lensed by a foreground elliptical galaxy (at the redshift $z_d$), multiple bright images of the active galactic nucleus (AGN) are formed together with the arcs of its host galaxy.
Time delays between multiple images can be measured from monitoring of variability of the AGN light curves. From theoretical point of view, lensing time delay is determined by both the geometry of the Universe (different paths of rays forming different images) as well as the Shapiro effects through \citep{1964PhRvL..13..789S}
\begin{equation}
\Delta t_{\rm AB} = D_{\Delta t}\left[\phi(\theta_{\rm A},\beta)-\phi(\theta_{\rm B},\beta)\right]=D_{\Delta t}\Delta\phi_{\rm AB}(\xi_{\rm lens}),
\end{equation}
where $\phi(\theta,\beta)=\left[{(\theta-\beta)^2}/{2}-\psi(\theta)\right]$ is the Fermat potential at images, $\beta$ is the source position, $\psi$ is   effective
lensing potential (the integral of the Weyl potential along the line-of-sight) obeying the Poisson equation $\nabla^2\psi=2\kappa$, where $\kappa$ (so called convergence) is the surface mass density of the lens in units of critical density $\Sigma_{\rm crit}=D^{A}_{\rm s}/(4\pi D^{A}_{\rm d}D^{A}_{\rm ds})$, and $\xi_{\rm lens}$ denotes the lens model parameters. The cosmological background is reflected in the so-called "time delay distance" $D_{\Delta t}=(1+z_{\rm d})\frac{D^{A}_{\rm d}D^{A}_{\rm s}}{D^{A}_{\rm ds}}$, where subscripts $d$ and $s$ stand for lens (deflector) and source, and superscript $A$ denotes the angular diameter distance. The key point here is that the Fermat potential difference $\Delta\phi_{\rm AB}(\xi_{\rm lens})$ can be reconstructed by high-resolution lensing imaging from space telescopes.

In the framework of parametrized post-Newtonian (PPN) formalism, the inferred lensing mass parameters are rescaled by a factor of $(1+\gp)/2$. Hence, we denote the actually inferred lens model parameters in the Fermat potential as $\xi^{'}_{\rm lens}$.
Adopting the notations in work \cite{2020MNRAS.497L..56Y}, one can write the time-delay distance as
\begin{equation}
D_{\Delta t}=(1+z_{\rm d})\frac{D^{A}_{\rm d}D^{A}_{\rm s}}{D^{A}_{\rm ds}}=\frac{\Delta t_{\rm AB}}{\Delta\phi_{\rm AB}(\xi^{'}_{\rm lens})}\,.
\label{eq:ddt}
\end{equation}
Thus, the time delay distance can be obtained from both the measurements of time delay and the actual reconstructed Fermat potential with parameter  $\xi^{'}_{\rm lens}$. The only difference in this equation is that in the case of $\gp \neq 1$ under the PPN framework  the inferred lens model parameters are $\xi^{'}_{\rm lens}$ but not the original  $\xi_{\rm lens}$ under GR.

On the other hand, assuming some explicit model of the lens such as the simplest Singular Isothermal Sphere (SIS) model (or its extensions like singular ellipsoid SIE or power-law model), observations regarding stellar kinematics (dynamical mass determination) allow to obtain the following distance ratio
\begin{equation}
\frac{D^{\rm A}_{\rm s}}{D^{\rm A}_{\rm ds}}=\frac{\sigma_v^2}{c^2J(\xi_{\rm{lens}}, \xi_{\rm{light}}, \beta_{\rm{ani}})}\;,
\label{eq:dsdds1}
\end{equation}
where $\sigma_v$ is the line of sight (LOS) projected stellar velocity dispersion
of the lens galaxy. This distance ratio provides a valuable extra constraint. 
The function $J$ captures all of the model components calculated from the lensed images and luminosity-weighted projected velocity dispersion (from the spectroscopy). Since the radial velocity dispersion $\sigma_v$ can be modelled via the anisotropic Jeans equation, arguments of $J$ function comprise lens model parameters $\xi_{\rm{lens}}$, parameters related to luminosity distribution of the lens
$\xi_{\rm{light}}$ and anisotropy parameter $\beta_{\rm{ani}}$. More details concerning modelling issues related to the function $J$ can be found in Section 4.6 of \citep{2019MNRAS.484.4726B}. Stellar kinematics of lensing galaxies is sensitive to the Newtonian potential only, which is independent of PPN parameter. Thus, the inferred distance ratio $D^{A}_s/D^{A}_{ds}$ combined with the well-measured velocity dispersion is independent of the cosmological model and time delays, but still relies on the lens model $\xi_{\rm{lens}}$ \citep{2016JCAP...08..020B,2019MNRAS.484.4726B}.
The lens model parameter in $J$ is the "unrescaled" $\xi_{\rm{lens}}$. If we replace  $\xi_{\rm{lens}}$  with $\xi^{'}_{\rm{lens}}$, the resulted distance ratio shall also be rescaled, correspondingly
\begin{equation}
\frac{2}{1+\gamma_{\rm{PPN}}}\frac{D^{\rm A}_{\rm s}}{D^{\rm A}_{\rm ds}}=\frac{\sigma_v^2}{c^2J(\xi'_{\rm{lens}}, \xi_{\rm{light}}, \beta_{\rm{ani}})}\;.
\label{eq:dsdds2}
\end{equation}

The PPN parameter is introduced here explicitly, and this formula was widely used in the works on constraining the GR parameter for strong gravitational lensing systems \cite{2017ApJ...835...92C,2022ApJ...927...28L,2022ApJ...927L...1W,2024MNRAS.528.1354L}.
Furthermore, we can define $D'^{A}_d=\frac{1+\gamma_{\rm{PPN}}}{2}D_d^A$, and by combining Eqs.~(\ref{eq:ddt}) and (\ref{eq:dsdds2}), the angular diameter distance to lens can be expressed as
\begin{eqnarray}
D'^{A}_{\rm d}=\frac{1}{1+z_{\rm d}}\frac{c\Delta t_{\rm AB}}{\Delta \phi_{\rm AB}(\xi'_{\rm{lens}})}\frac{c^2J(\xi'_{\rm{lens}},\xi_{\rm{light}},\beta_{\rm{ani}})}{\sigma_v^2}\,.
\label{eq:ddp}
\end{eqnarray}
Let us stress that this distance is unaffected by cosmic opacity\cfootnote[black]{In the case of gravitational lensing observables,  cosmic opacity can change the absolute intensity (magnitude) of images but not the relative intensity, thus not biasing the distance determination in strong lensing system. Besides, the velocity dispersion based on spectroscopic measurements are also unaffetced by the opacity.}.
In fact, one could use the time delay distance (Eq. \ref{eq:ddt}) to test gravity theory, but this will inevitably introduce cosmic curvature parameter, because $D^A_{\rm ds}$ is not directly observable. More importantly, one of the main of obstacle for lensing mass modelling is the mass-sheet degeneracy, which is completely circumvented by testing gravity using angular diameter distances.
Many factors such as the Hubble constant, dark energy model and cosmic opacity (e.g. caused by light extinction by intergalactic dust) affect determination of cosmological distances. This would bias most straightforward tests of GR and reliable test of GR should be pure enough to ensure its validity and being unbiased. Therefore, we point out in this work that GR testing by gravitational lensing could be independent of these factors.

The current and future programs for time-delay with lensed quasars have great progress, such as H0LiCOW \cite{2020MNRAS.498.1420W}, COSMOGRAIL \cite{2005A&A...436...25E}, STRIDES \cite{2018MNRAS.481.1041T}, and SHARP, which have now combined in the TDCOSMO collaboration \cite{2021A&A...649A..61B,2023A&A...673A...9S}.
Although our proposed method has many advantages, yet the existing sample of time-delay lenses is still insufficient, so we turn to a new generation of wide and deep sky observations and make appropriate simulations.
Regarding the assessment of the uncertainties of $D^A_d$, for each lens one should followed the process presented in \citep{2020MNRAS.493.4783Y}, which resulted in 1.8\% precision of $D_d^A$ for RXJ1131-like system.
In a realistic situation, the analysis of each lens is very complicated, and such simulations are beyond this work.
Instead, we follow the work of \citep{2016JCAP...04..031J} and consider that the main sources of uncertainties give the expected uncertainty level for $D^A_d$ determination. These uncertainties, including the measurement of time delays, the mass fluctuation along the LOS, recovering  stellar kinematics from measured velocity dispersions, the lens mass parameterisation from highly resolved imaging will result in a few percent level of $D_d^A$ determination accuracy.
We take the  5\% uncertainty on the measurement of $D_d^A$ as the best case scenario, which was also
also used in \citep{2016JCAP...04..031J,2011PhRvD..84l3529L} and as the aim of the H0LiCOW program \citep{2017MNRAS.468.2590S}. In the conservative scenario, we also adopt 10\% uncertainty on $D_d^A$ for comparison.  This conservative uncertainty is estimated from the \citep{2020MNRAS.494.6072S}, which used
simulations to obtain realistic error estimates with current/upcoming instruments, e.g. OH Suppressing InfraRed Imaging Spectrograph on Keck \citep{2006NewAR..50..362L}, and Near-Inrared Spectrograph on \textit{James Webb Space Telescope} \citep{2022A&A...661A..83B}, and InfraRed Imaging Spectrograph on the Thirty Metre Telescope \citep{2016SPIE.9909E..05W}.

\begin{figure}
\centering
\includegraphics[width=0.9\linewidth]{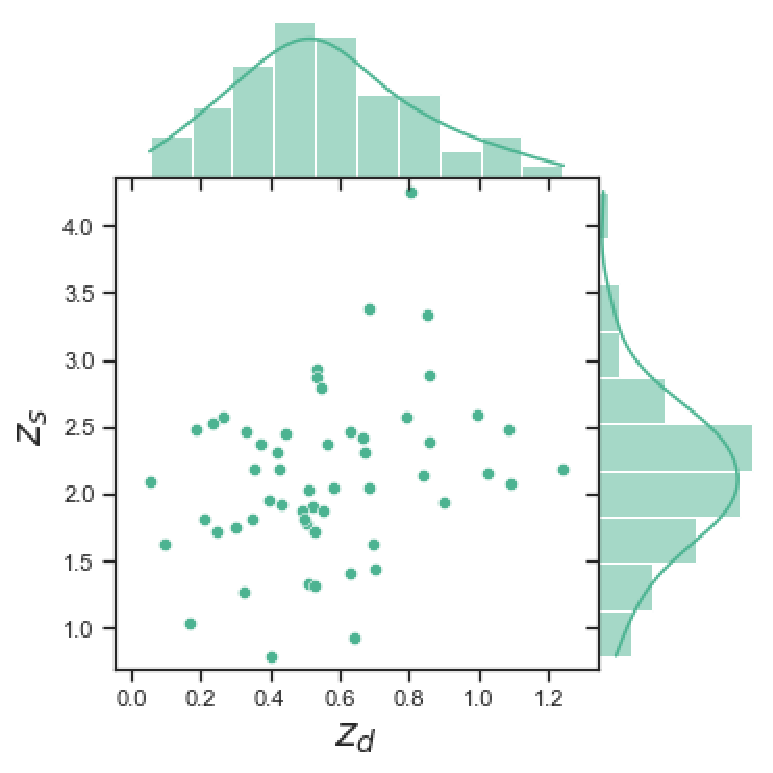}
\caption{ Redshift distribution of lenses and sources in a sample of 55 strong lensing systems expected to be observed in current and
upcoming projects.}\label{fig1}
\end{figure}

In the near future, the next generation of wide and deep sky surveys, with improved depth, area and resolution, may increase the current galactic-scale lens sample sizes by orders of magnitude.
For example, the upcoming Legacy Survey of Space and Time (LSST) conducted in Vera Rubin Observatory, is expected to find ten thousand lensed quasars \citep{2019ApJ...873..111I}. However, considering that the metric efficiency is about 20\%, the Time Delay Challenge (TDC) program showed that only 400 well-measured time delays are available \citep{2015ApJ...800...11L}. More precisely, LSST will discover 10000 lensed quasars, but it has been estimated that only 2000 lensed quasar systems would have a long enough (10$<\Delta t <$ 120 days) time delays measurements. The TDC expects to be able to make time delay measurements with the high accuracy in at least 20\% efficiency of an LSST sample of 2000 lenses. This would correspond to a well-measured sample of around 400 lensed quasars. In this simulation, we use the OM10 catalog of mock lenses\footnote{https://github.com/drphilmarshall/OM10} \citep{2010MNRAS.405.2579O}, which has the distribution of time-delay lenses expected from the LSST. Besides, one would need auxiliary data comprising  high resolution imaging and spectroscopy from instruments such as the Hubble Space Telescope (HST) and ground-based observatories, in order to have all ingredients needed for accurate lens mass modeling and determination of $D_d^A$. It should be pointed out that LSST will discover a number of fainter, smaller-separation lenses where it is not clear that the same level of precision of lens model reconstruction can be reached.
Therefore, we follow the work of \citep{2016JCAP...04..031J} and set the following criteria: 1) the lensed quasar image separation should be greater than $1$ \textit{arcsec}; 2) the third brightest quasar image has i-band magnitude $m_{i}<21$ \textit{mag}; 3) the lens galaxy has $m_{i}<22$ \textit{mag}; (4) the lensed quasar image is quadruple imaging lens systems (this type of system provide more information to break the Source-Position Transformation \citep{2013A&A...559A..37S,2014A&A...564A.103S}).
After applying these conditions to the OM10 catalog, 55 high-quality quadruple lens systems are kept \citep{2016JCAP...04..031J}.
Fig. \ref{fig1} shows the source and the lens redshift distributions of 55 quadruple lenses that match the selection criteria.

\subsection{Luminosity distance from gravitational waves}

From Eq. (\ref{eq:ddp}), it is clear that the PPN parameter and the angular diameter distance are entangled. 
In order to break this degeneracy and complete the test of the gravity theory, one needs independent, complementary distance measurements. For this purpose, we turn to standard sirens.

Simultaneous detections of GW signal \citep{Abbott17} from the binary neutron star (NS-NS) merger and the electromagnetic (EM) counterparts \citep{2017ApJ...848L..12A} from the same transient source opened a new era of multi-messenger astronomy.  GW signals from coalescing compact binary systems, i.e. binary black holes (BH-BH) \citep{Abbott16}, neutron stars (NS-NS) or possible mixed neutron star - black hole (NS-BHs) systems provide us direct measurements of luminosity distances $D^L$. Hence, they are called standard sirens \citep{Schutz1986}. Standard sirens are self-calibrating, meaning that the luminosity distances can be directly inferred from the detected waveforms using matched-filter method. Redshift of the source $z$ is unfortunately non-measurable in GW domain alone, and one has to identify the host galaxy in order to obtain $z$.
There are two advantages of standard sirens: firstly, GW signals are unaffected by cosmic opacity, and  propagate through the Universe without any absorption and dissipation. Secondly, GW signals from standard sirens provide direct luminosity distance measurements instead of the relative distances as in case of SNe Ia. Thus $D^L$ from standard sirens do not need to be anchored to the cosmic distance ladder.

Einstein Telescope (ET)\cfootnote[black]{ The Einstein Telescope Project, \url{https://www.et-gw.eu/et/}.} \citep{Dalal2006,Taylor12} is the third-generation gravitational wave detector, which is designed to have a fantastic sensitivity in the frequency range of $1-10^4$ Hz. See \citep{Abernathy11} for the details of the ET conceptual design study.  We simulate GW signals based on the foreseen performance of the ET. Compared with the advanced detectors such as AdLIGO and AdVirgo, such proposed third-generation detector aims for a broadband factor of 10 sensitivity improvement, especially for the characteristic distance sensitivity \citep{Taylor12b}. In this simulation, we follow the work by \citep{2023MNRAS.518.3372C,2023PhRvD.107l3519C}, and assume that the GW sources are caused by binary merger of the NS with the NS, which can generate the intense short gamma ray bursts (SGRBs) with measurable source redshifts, or binary merger of the BH with the BH, which the redshift information comes from a statistical analysis over a catalogue of potential host galaxies.

Following the strategy proposed in \citep{2023MNRAS.518.3372C,2023PhRvD.107l3519C}, the uncertainty for luminosity distance obtained from the GW signal from merger event is
\begin{equation}
\sigma^2_{\rm tot}= \sigma^{2}_{\rm inst}+ \sigma^{2}_{\rm
len}+\sigma_{pec}^2,
\end{equation}
where $\sigma_{\rm inst}$ is the instrumental measurement uncertainty and calculated by
\begin{equation}
\sigma_{\rm inst}\simeq\sqrt{\bigg\langle\frac{\partial \mathcal{H}(f)}{\partial D^{L,GW}}, \frac{\partial \mathcal{H}(f)}{\partial D^{L,GW}}\bigg\rangle^{-1}},
\end{equation}
where $\mathcal{H}(f)$ is the Fourier transform of \textit{strain} $h(t)$ of a chirp waveform in frequency domain. Since the \textit{strain} $h$ and $\mathcal{H}$ is inversely proportional to $D^{L,GW}$, we obtain $\sigma_{\rm inst}\simeq D^{L,GW}/\rho$, where $\rho$ denotes the combined  signal-noise-ratio (SNR) for the network of independent interferometers, determined by the square root of the inner product of $\mathcal{H}(f)$.  Considering the uncertainty from the inclination angle $\iota$ would also affect the SNR,  the maximal inclination angle effect from $\iota=0^\circ$ to $\iota=90^\circ$ on the SNR, we choose to double the instrumental uncertainty of $D^L$ as the upper limit, as was proposed in the recent study of the third-generation GW detector (ET) based on Fisher information matrix \citep{Cai17}, i.e.,
\begin{equation}
\sigma_{\rm inst}\simeq \frac{2D^{L,GW}}{\rho}.
\end{equation}
The second uncertainty term $\sigma_{\rm len}$ is the weak gravitational lensing effect caused by the large scale structure, and ignoring it would result in the biased distance estimation. Following the work \citep{2023MNRAS.518.3372C,2023PhRvD.107l3519C},   this uncertainty is modeled as \cite{2010PhRvD..81l4046H,2016JCAP...04..002T}
\begin{equation}
\sigma_{lens}=0.066\left(\frac{1-(1+z)^{-0.25}}{0.25}\right)^{1.8}D^L(z)F_{delens}(z),
\end{equation}
where $F_{delens}(z)= 1- \frac{0.3}{\pi /2}\arctan{{z}/{z_*}}$, with $z_*=0.073$
\citep{2016JCAP...04..002T}. The latter factor takes into account the possibility to reduce the uncertainty due to weak lensing with the future detectors such as the Extremely Large Telescope \citep{2021PhRvD.103h3526S}.
The final uncertainty term $\sigma_{pec}$ is caused by the peculiar velocity of the host galaxy  \citep{2017ApJ...848L..31H,2020MNRAS.492.3803H,2021A&A...646A..65M}, and  can be approximated by a fitting formula \citep{2006ApJ...637...27K}
\begin{equation}
\sigma_{pec}=\left[ 1+\frac{c(1+z)^2}{H(z)D^L (z)}\right]\frac{\sqrt{\langle v^2\rangle}}{c}D^L (z)\,,
\end{equation}
where the averaged peculiar velocity $\langle v^2\rangle$  is set as  $500$ km/s, in agreement with the observed values in galaxy catalogs \citep{2000ApJ...538...83C}.

Although the next generation ground-based detector ET is expected to detect tens or hundreds of thousand NS-NS inspiral events per year up to the redshift $z \sim 2$ and NS-BH mergers up to $z \sim 5$ with $\rho>8$, yet
 SGRBs necessary for $z$ measurements are strongly beamed, and only the nearly face-on configurations of NS-NS or NS-BH mergers are useful. Probability of their occurence is $\sim 10^{-3}$.
Assuming that the redshift distribution of simulated GW sources follows the cosmic star formation history \citep{2001MNRAS.324..797S}, we sample the the mass of neutron star
and black hole within [1, 2] $M_{\odot}$ and [3, 10] $M_{\odot}$, respectively.
Thus, we simulate 1000 GW events observable in the ET and their accurate redshifts up to $z_{GW}=5$. Matching these events with the redshifts of the lens $z_d$ needed for testing the PPN parameter, reduces the sample to 300 sources within redshift $z_{GW}<1.25$.

\section{Results and discussion} \label{sec:method}
\begin{table}
\renewcommand\arraystretch{1.8}
\caption{\label{tab:result} Summary of the constraints on the $\gp$ parameter using the combination of strong lensing systems plus GW events.}
\begin{tabular}{l| c| c |c}
\hline
\hline
$\sigma_{D^A}/{D^A}=5\%$  & $\rm\gamma_0$ &$\rm\gamma_1$ &$\rm\gamma_2$  
\\
\hline
$\gp=\gamma_0$  & $1.003^{+0.018}_{-0.018}$ & $-$ & $-$ \\
\hline
$\gp=1+\gamma_1*z$  & $-$ & $0.006^{+0.030}_{-0.031}$ & $-$ \\
\hline
$\gp=1+\gamma_2*a$  & $-$ & $-$ & $0.008^{+0.028}_{-0.028}$ \\
\hline
\hline
$\sigma_{D^A}/{D^A}=10\%$  & $\rm\gamma_0$ &$\rm\gamma_1$ &$\rm\gamma_2$
\\
\hline
$\gp=\gamma_0$  & $1.003^{+0.032}_{-0.033}$ & $-$ & $-$ \\
\hline
$\gp=1+\gamma_1*z$  & $-$ & $0.007^{+0.050}_{-0.050}$ & $-$ \\
\hline
$\gp=1+\gamma_2*a$  & $-$ & $-$ & $0.004^{+0.050}_{-0.049}$ \\
\hline
\hline
\end{tabular}
\end{table}

\begin{figure*}
\centering
\includegraphics[width=0.32\linewidth]{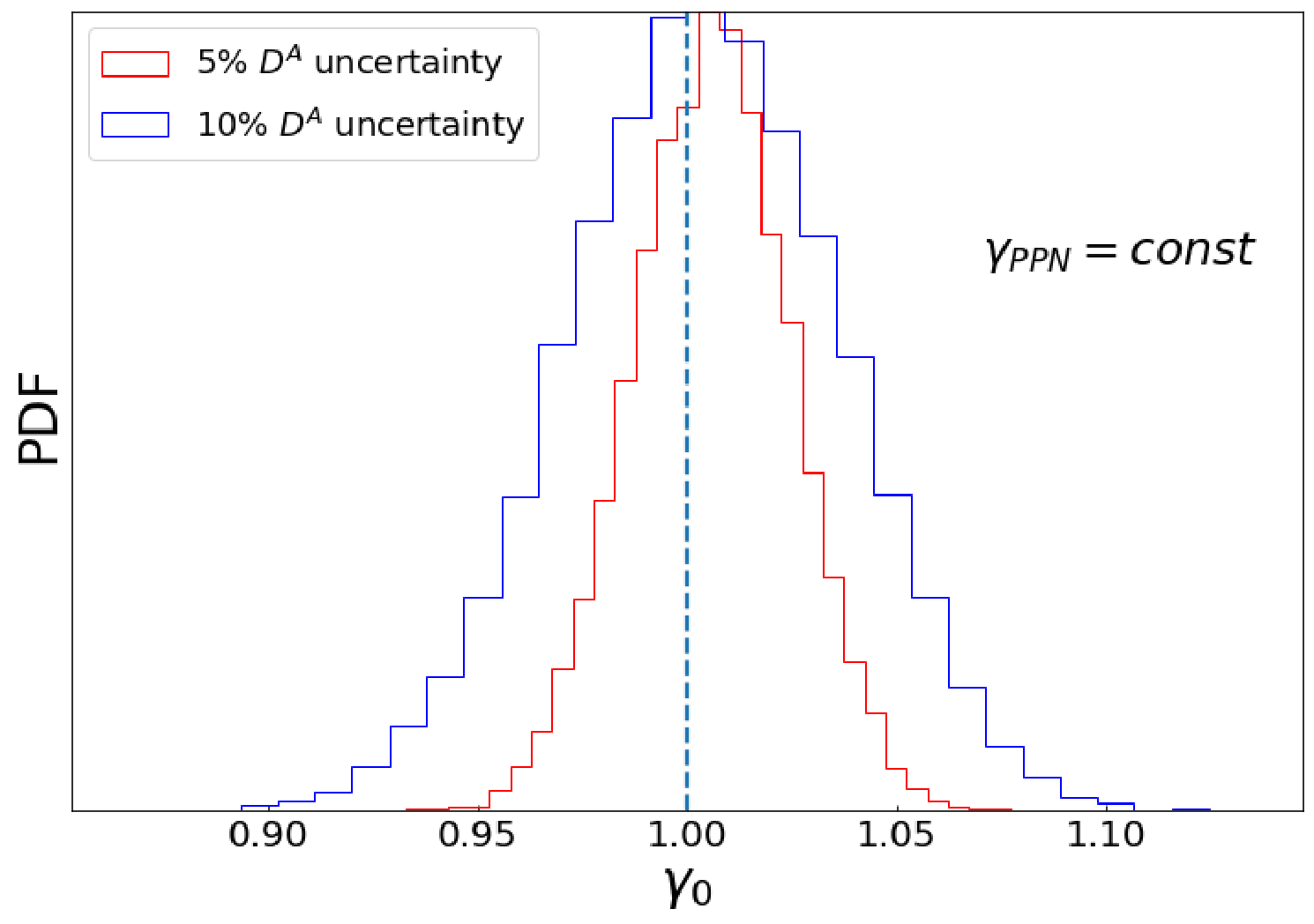}
\includegraphics[width=0.32\linewidth]{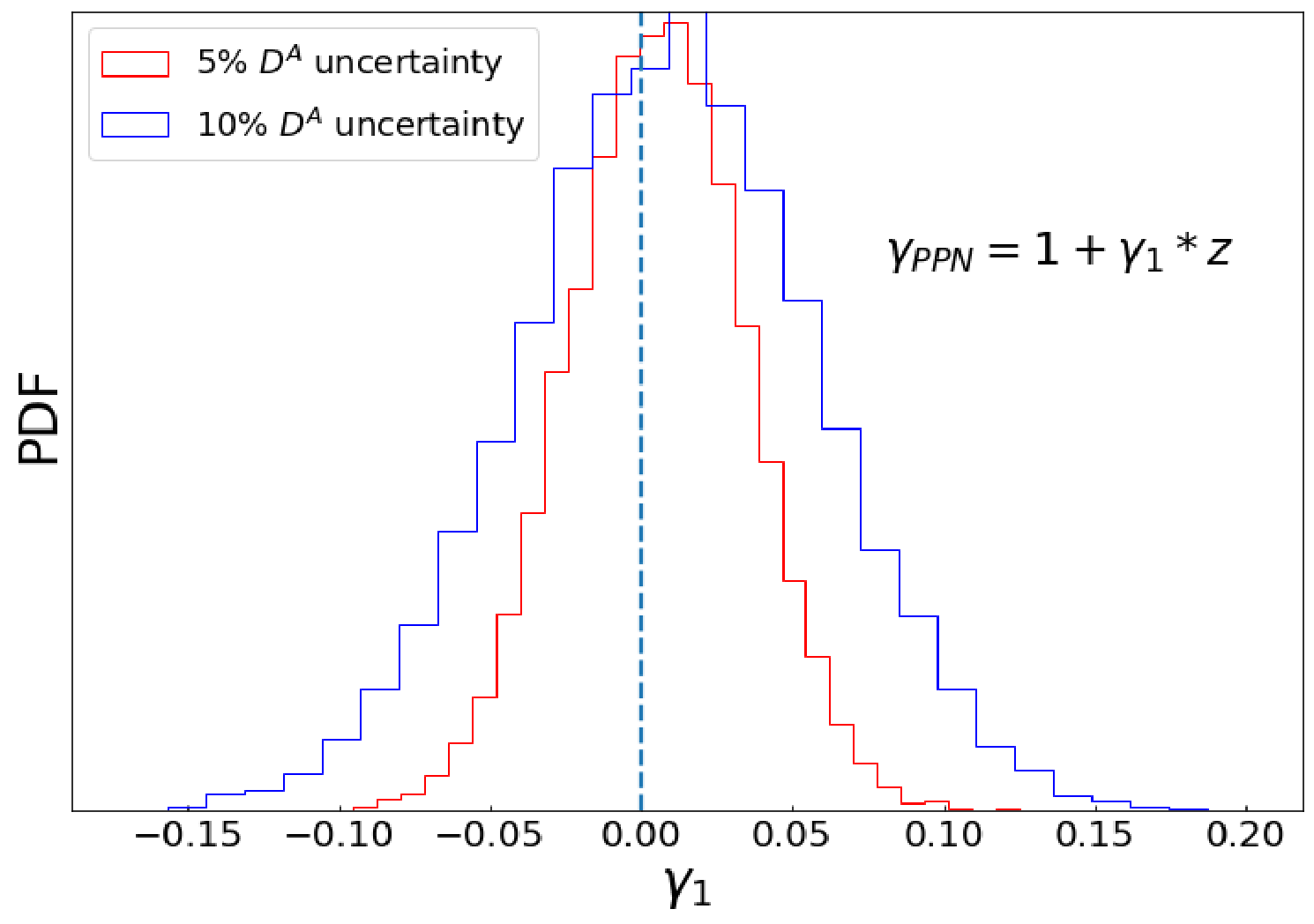}
\includegraphics[width=0.32\linewidth]{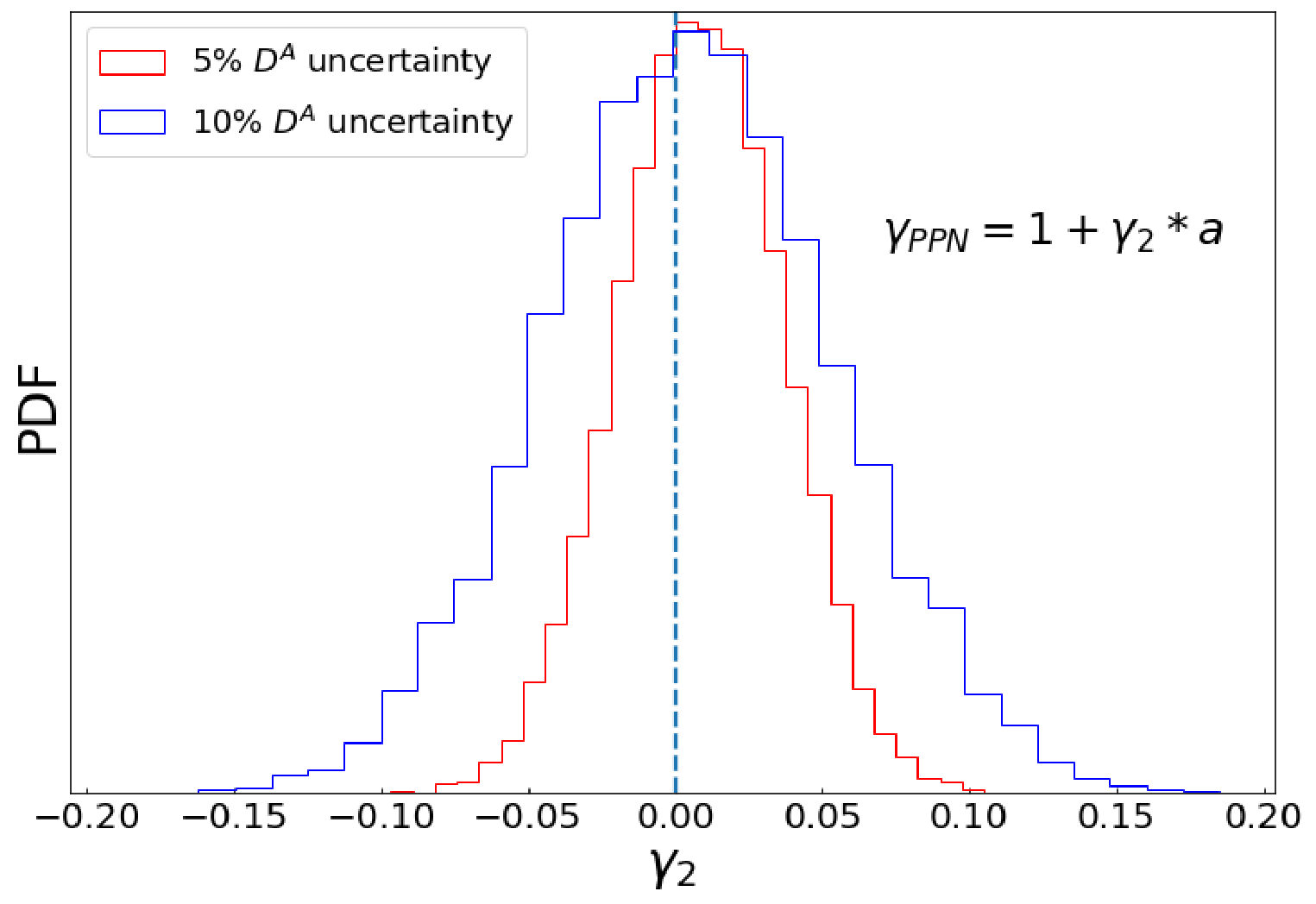}
\caption{ Posterior distributions for PPN parameters obtained by the combination of the future LSST survey and ET detector yields.
Three panels illustrate the following parametrizations: $\gp=\gamma_0$ constant case (left), $\gp=1+\gamma_1*z$ varying with redshift (middle), and $\gp=1+\gamma_1*a$ varying with the scale factor (right) .}
\label{fig4}
\end{figure*}
To demonstrate the performance of our method to detect the possible deviation of GR, we use three parameterized forms of $\gp$: 1) $\gp=\gamma_0$  being constant; 2) $\gp=1+\gamma_1*z$ changing with redshift; 3) $\gp=1+\gamma_2*a$ changing with the scale factor $a=1/(1+z)$, which unlike the redshift is the gravitational degree of freedom. The above mentioned PPN parameters and their uncertainties are fitted by minimizing the $\chi^2$ objective function defined in the following way:
\begin{equation}
\chi^2=\sum^{55}_{i=1}\bigg[\frac{(D'^{A}_{d, i}-D^L_{GW,i}(1+z)^{-2})^2}{\sigma_{D^L_{GW,i}}^2+\sigma_{D'^{A}_{d,i}}^2}\bigg],
\end{equation}
{where $\gp$ parameter is implicitly present in $D'^{A}_{d}$. The total uncertainty consists of observational uncertainties from GW and strong lensing contributions (see Section 2. B and C). The uncertainty $\sigma_{D^L_{GW}}=\sigma_{\rm tot}/(1+z)^{2}$ is calculated using $\sigma_{\rm tot}$ given by Eq. (7), strong lensing related uncertainties are given by $\sigma_{D'^{A}_{d}}=5\%D'^{A}_{d}$ and $\sigma_{D'^{A}_{d}}=10\%D'^{A}_{d}$ for the best case scenario and the conservative scenario, respectively.  We simulated ten thousand realizations of the data with different random seeds and repeated the minimization process to get an unbiased estimation for the PPN parameters.  We emphasize that in this work we do not attempt to constrain PPN parameter from real data, but propose an unbiased approach and discuss its ability to explore the presence of possible evolution or deviations from GR on simulated data based on future observational forecasts.

The final Probability Distribution Functions (PDFs) for $\gp$ are reported in Table \ref{tab:result} and displayed in Fig. \ref{fig4}.
Since the PDFs are approximately Gaussian-like, we calculate the standard deviations as the $1\sigma$ uncertainty levels. For the 5\%  uncertainty (optimistic case) on the $D_d^A$ measurements from strong lensing systems, our results regarding $\gp=\gamma_0$ constant case suggest that the uncertainty on $\gamma_0$ is 0.018, corresponding to 1.8\% precision.
The precision obtained by our method is $>10$  times better than the one obtained from current four well-measured H0LiCOW quasars lenses, and almost 3 times better that the one from simulated future 40 lensed quasar systems \citep{2020MNRAS.497L..56Y}. Meanwhile, the precision obtained by our method is $>4$ times better than expected from the simulated future 10 strongly lensed fast radio bursts (FRBs) systems \citep{2022MNRAS.516.1977G}.  This result demonstrates the superiority of our method. There are two reasons for this. Firstly, only one parameter $\gamma_0$ is estimated in this work, whereas the previous work was simultaneously estimating $H_0$ and PPN parameters, and the distance information was derived there by assuming a concrete cosmological model. This affects the ability to constrain PPN parameter alone. Secondly,  one of the main of obstacle for lensing mass modelling is the mass-sheet degeneracy\cfootnote[black]{It is a mathematical degeneracy that leaves the lensing observables unchanged, while rescaling the absolute time delay, and thus the inferred $H_0$.}, which is considered the dominant source of residual modeling error in time-delay cosmography. Our method circumvents the mass-sheet degeneracy by using angular diameter distances to perform the test of gravity,  which greatly reduces systemic uncertainty and further improves the precision on PPN parameter in our work.

Some modified gravitational theories, such as the nonlocal Gauss-Bonnet gravity \citep{2019PhRvD..99f4044T} require that PPN parameter may evolve with redshift or the scale factor. With this motivation, we further assume a parameterized form of $\gp$ changing with redshift or the scale factor. The results are also shown in Fig. \ref{fig4} and reported in Table \ref{tab:result}.
We obtain $\Delta \gamma_1=0.030$ and $\Delta \gamma_2=0.028$ in the optimistic scenario. Compared with PPN parameter treated as a constant, the precision of PPN parameters evolving with redshift or the scale factor is noticeably reduced. In Fig. \ref{fig5}, we plot the reconstruction of $\gp$ for the different parameterizations with their uncertainties, and a prediction from a nonlocal Gauss-Bonnet theory with specific parameters \citep{2009PhLB..671..193C}. It is clearly seen that if GR does not hold and is replaced by this nonlocal  gravity, our method will distinguish them with very high precision. Here, for the purple dashed line in Fig. \ref{fig5}, we adopt the conventions in \citep{2019PhRvD..99f4044T} and set the model parameter $\alpha=5.6\times10^{-6}c^4/H_0^4$, the integral constant $C_1=0$, and the matter density $\rho\propto a^{-3}$. Note that, generally nonlocal Gauss-Bonnet gravity predicts $|1-\gp|=\mathcal{O}(1)$ \citep{2019PhRvD..99f4044T}. The $\alpha$ we use here is too small to explain the cosmic late-time acceleration, which is the initial motivation to propose this theory \citep{2009PhLB..671..193C}. Therefore, this purple line corresponds to a toy model. This illustration is just to show that $\gp$ in modified gravity can be very close to 1 at low redshifts and deviate from 1 at high redshifts. Such possible behavior of $\gp$ highlights the importance of constraining it at high redshifts. In addition, the method proposed in this paper is independent of any specific cosmological models and gravitational theories.

The main purpose of this work is to quantify the ability of upcoming TDSL plus GW to test GR against modified gravity theories.  The key question now arises: Is this combination of observations sufficient to detect possible effects of deviation from GR? They can hardly become competitive to high-precision measurements of $\gp=1+(2.1\pm2.3)\times10^{-5}$ in the solar system made by Cassini mission.
Considering the 10\% uncertainty (conservative case) on the $D_d^A$ measurements from strong lensing systems, the precision on PPN parameter is almost half of that for the optimistic case. This suggests that the main source of uncertainty in our method is the observational error from the angular diameter distance measured by the gravitational lensing system. This is reasonable, considering that only 55 gravitational lensing systems have been used in this work, and the uncertainty of the PPN parameters caused by systematic errors is much greater than that caused by statistical errors. If we take a more optimistic estimation presented by   \citep{2020MNRAS.493.4783Y}, the future uncertainty on $D_d^A$ can even decrease dramatically to the order of 1.8\%, which indicates a more promising future of TDSL plus GW technique for testing general relativity.
With the upcoming sky surveys such as LSST, the number of lensing systems will increase dramatically. Hence, one can expect that our approach will greatly improve the measurement precision on $\gp$ at the galactic scale.

\begin{figure}
\begin{center}
\includegraphics[width=0.9\linewidth]{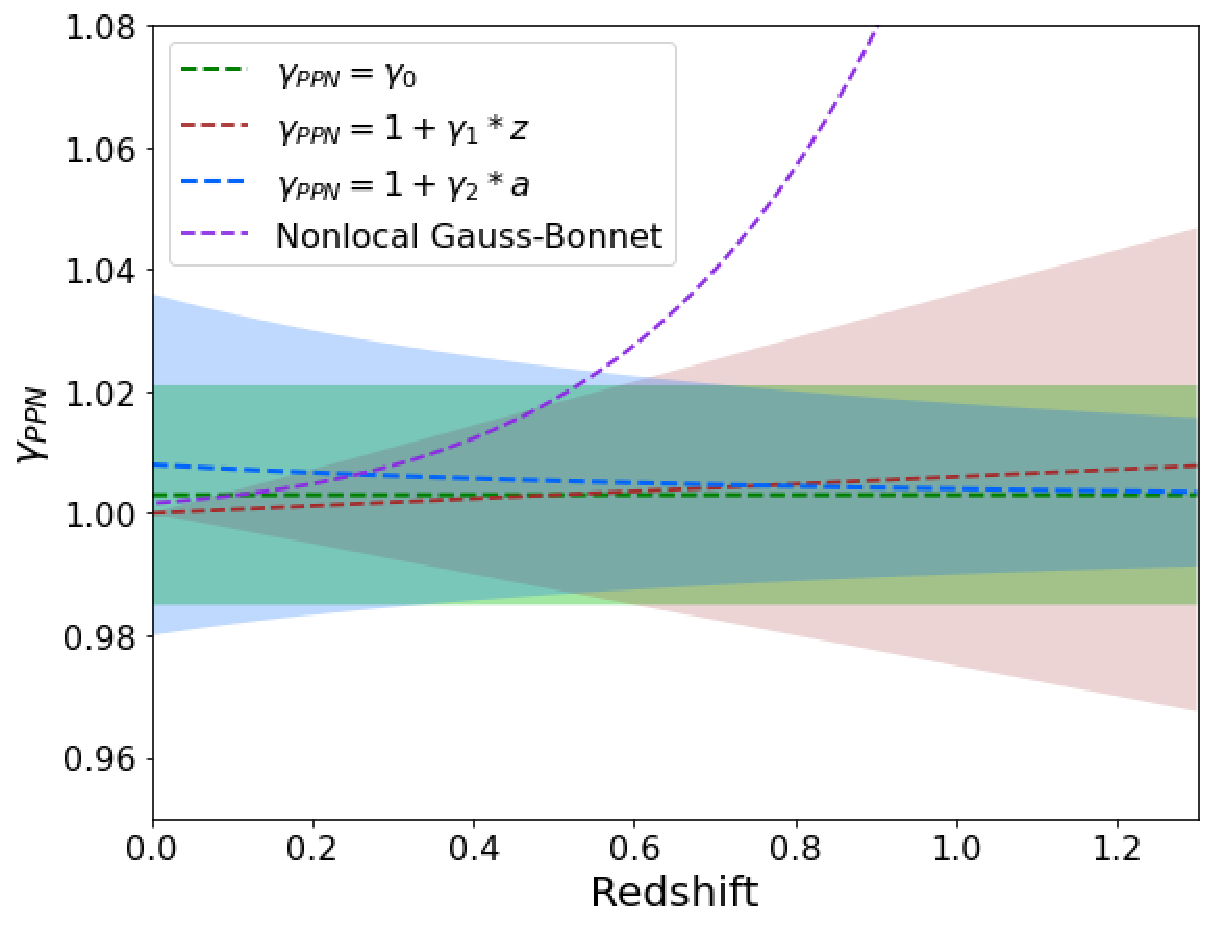}
\end{center}
\caption{The reconstructed $\gp$ for the three parameterizations with their  $1\sigma$ uncertainties. The purple dashed line represents the nonlocal Gauss-Bonnet theory. The green, pink, and blue bands correspond to the $1\sigma$ regions of the best-fitting in the constant, evolution with the redshift, and evolution with the scale factor  cases, respectively.  }
\label{fig5}
\end{figure}

\section{Conclusion}
{In this work, we investigated, for the first time, the possibility of estimating post-Newtonian parameter $\gp$ by combining well measured strongly lensed systems with gravitational wave signals. The combination of strong lensing and self calibrating standard sirens observed in GWs enables to avoid the possible bias that comes with assuming a particular cosmological model.} Meanwhile, such combination provides a relatively pure and unbiased method for GR testing  at the galactic scales and high redshifts, and brings various benefits. It can reduce possible bias on GR testing induced by cosmic curvature, cosmic opacity, the dark energy, the Hubble constant (the $D^A_d$ provided by strong lensing and the $D^L$ provided by gravitational waves are both absolute distances and are therefore unaffected by the Hubble constant), and the mass-sheet degeneracy inside the lens. Combining measurements of time delay, stellar velocity dispersion, high resolution images, LOS environment modeling from upcoming LSST survey and GWs acting standard siren from future ET detector, we obtained the precision of $\sim$ 1.8\% (optimistic case) and 3.2\% (conservative case) in the case of $\gp$ being a constant. Secondly, we considered the case of $\gp$ parameter displaying evolution with redshift or the scale factor and studied the precision with which this effect can be revealed. {This combination of data also allows us to infer post-Newtonian parameter and detect possible deviation of GR at different redshifts.}
Although the precision of testing GR at the galactic scale can hardly achieve the results competitive to the measurements within the solar system, yet our technique would be able to distinguish the departures from the GR quite precisely. We also indicated a more promising future of TDSLs plus GWs for testing general relativity.

As a final remark, there are many potential ways to improve our method. For instance, current and future surveys like the Dark Energy Survey (DES) \citep{2018MNRAS.481.1041T}, the Hyper SuprimeCam Survey \citep{2017MNRAS.465.2411M}, and the 
Legacy Survey of Space and Time (LSST)
\citep{2010MNRAS.405.2579O,2015ApJ...811...20C} will bring us hundreds of thousands of lensed quasars in the most optimistic discovery scenario.
Even if only some small fraction of them will have precise measurements of time delays between multiple images, the resulting statistics will outshine current catalogs. With high-quality auxiliary observations, one can use high-cadence, high-resolution and multi-filter imaging of the resolved lensed images, to derive an accurate determination of the Fermat potential, which will increase the precision of time delay distance by an order of magnitude.  On the other hand, the first strongly lensed supernova ``SN Refsdal" \cite{2015Sci...347.1123K} with multiple images opened a new window for astrophysics and cosmology with the concept of ``lensed transients".
We also expect that the discovery of various lensed transients explosive sources like gamma ray bursts (GRBs), fast radio bursts (FRBs) and even GWs  will give us new advantages over the traditional targets in studying the Universe \cite{2022ChPhL..39k9801L,2017NatCo...8.1148L}.
At the same time, we also expect in the future, the synergies between GW and EMW observations in various bands will yield reliable cosmological probe. It is reasonable to expect that our approach will play an increasingly important role in precise testing the validity of general relativity.

\acknowledgments
This work was supported by National Natural Science Foundation of China under Grant Nos. 12222302, 12203009, 11973034. Tonghua Liu was supported by Chutian Scholars Program in Hubei Province (X2023007);  Kai Liao was supported by Funds for the Central Universities (Wuhan University 1302/600460081). Shuxun Tian was supported by the Fundamental Research Funds for the Central Universities (Beijing Normal University 10800-310400209522). Marek Biesiada was  supported by Hubei Province Foreign Expert Project (2023DJC040).


\begin{thebibliography}{}
\expandafter\ifx\csname natexlab\endcsname\relax\def\natexlab#1{#1}\fi
\providecommand{\url}[1]{\href{#1}{#1}}
\providecommand{\dodoi}[1]{doi:~\href{http://doi.org/#1}{\nolinkurl{#1}}}
\providecommand{\doeprint}[1]{\href{http://ascl.net/#1}{\nolinkurl{http://ascl.net/#1}}}
\providecommand{\doarXiv}[1]{\href{https://arxiv.org/abs/#1}{\nolinkurl{https://arxiv.org/abs/#1}}}

\bibitem[Riess et al.(1998)]{1998AJ....116.1009R} Riess, A.~G., Filippenko, A.~V., Challis, P., et al.\ 1998, \aj, 116, 1009. 
\bibitem[Perlmutter et al.(1999)]{1999ApJ...517..565P} Perlmutter, S., Aldering, G., Goldhaber, G., et al.\ 1999, \apj, 517, 565. 
\bibitem[Wald(1984)]{1984ucp..book.....W} Wald R. \ 1984 "General Relativity."
\bibitem[Weinberg(1972)]{1972gcpa.book.....W} Weinberg, S.\ 1972, Gravitation and Cosmology: Principles and Applications of the General Theory of Relativity, by Steven Weinberg, pp. 688. ISBN 0-471-92567-5. Wiley-VCH , July 1972., 688
\bibitem[Planck Collaboration et al.(2016)]{2016A&A...594A..13P} Planck Collaboration, Ade, P.~A.~R., Aghanim, N., et al.\ 2016, \aap, 594, A13. 
\bibitem[Planck Collaboration et al.(2020)]{2020A&A...641A...6P} Planck Collaboration, Aghanim, N., Akrami, Y., et al.\ 2020, \aap, 641, A6. 
\bibitem[Riess et al. (2021)]{SH0ES} Riess, A.G., Casertano, S., Yuan, W., Bowers, J.B., Macri, L., Zinn, J.C. and Scolnic, D.\ 2021, \apj 908, L6.
\bibitem[Wong et al.(2020)]{2020MNRAS.498.1420W} Wong, K.~C., Suyu, S.~H., Chen, G.~C.-F., et al.\ 2020, \mnras, 498, 1420.

 \bibitem[Di Valentino et al.(2020)]{2020NatAs...4..196D} Di Valentino, E., Melchiorri, A., \& Silk, J.\ 2020, Nature Astronomy, 4, 196.
 \bibitem[Handley(2021)]{2021PhRvD.103d1301H} Handley, W.\ 2021, \prd, 103, L041301.
\bibitem[Abdalla et al.(2022)]{2022JHEAp..34...49A} Abdalla, E., Abell{\'a}n, G.~F., Aboubrahim, A., et al.\ 2022, Journal of High Energy Astrophysics, 34, 49. 
\bibitem[Weinberg(1989)]{Weinberg89} Weinberg, S. 1989, Rev. Modern Phys., 61, 1
\bibitem[Ashby(2002)]{Ashby02} Ashby, N. 2002, Phys. Today, 55, 41
\bibitem[Bertotti et al.(2003)]{Bertotti03} Bertotti, B., Iess, L., \& Tortora, P. 2003, Nature, 425, 374
\bibitem[Weisberg \& Huang(2016)]{2016ApJ...829...55W} Weisberg, J.~M. \& Huang, Y.\ 2016, \apj, 829, 55.
\bibitem[Shao(2021)]{2021PhyOJ..14..173S} Shao, L.\ 2021, Physics Online Journal, 14, 173. 
\bibitem[Isi et al.(2019)]{2019PhRvL.123l1101I} Isi, M., Chatziioannou, K., \& Farr, W.~M.\ 2019, \prl, 123, 121101. 
\bibitem[Ding et al.(2021)]{2021ApJ...921L..19D} Ding, H., Deller, A.~T., Fonseca, E., et al.\ 2021, \apjl, 921, L19.
 \bibitem[Gair et al.(2013)]{2013LRR....16....7G} Gair, J.~R., Vallisneri, M., Larson, S.~L., et al.\ 2013, Living Reviews in Relativity, 16, 7.
 \bibitem[Collett et al.(2018)]{Collett2018.Science.360.1342} Collett, T.~E., Oldham, L.~J., Smith, R.~J., et al.\ 2018, Science, 360, 1342. 
\bibitem[Will(2014)]{Will2014} Will, C. M. 2014, Living Reviews in Relativity, 17, 4

\bibitem[Ishak(2019)]{2019LRR....22....1I} Ishak, M.\ 2019, Living Reviews in Relativity, 22, 1. 
\bibitem[Uzan(2010)]{2010GReGr..42.2219U} Uzan, J.-P.\ 2010, General Relativity and Gravitation, 42, 2219. 
\bibitem[Koyama(2016)]{Koyama2016} Koyama, K. 2016, Rept. Prog. Phys., 79, 046902
\bibitem[Schutz (1986)]{Schutz1986}Schutz, B. F., 1986, Nature, 323, 310
\bibitem[Hu \& Sawicki(2007)]{Hu2007.PRD.76.064004} Hu, W. \& Sawicki, I.\ 2007, \prd, 76, 064004. 
\bibitem[Saaidi et al.(2011)]{Saaidi2011.PRD.83.104019} Saaidi, K., Mohammadi, A., \& Sheikhahmadi, H.\ 2011, \prd, 83, 104019. 
\bibitem[Thomas et al.(2023)]{Thomas2023.JCAP.04.016} Thomas, D.~B., Clifton, T., \& Anton, T.\ 2023, \jcap, 2023, 016. 
\bibitem[Tian \& Zhu(2019)]{2019PhRvD..99f4044T} Tian, S.~X. \& Zhu, Z.-H.\ 2019, \prd, 99, 064044. 
\bibitem[Shapiro(1964)]{1964PhRvL..13..789S} Shapiro, I.~I.\ 1964, \prl, 13, 789. 

\bibitem[Yang et al.(2020)]{2020MNRAS.497L..56Y} Yang, T., Birrer, S., \& Hu, B.\ 2020, \mnras, 497, L56.
\bibitem[Birrer et al.(2019)]{2019MNRAS.484.4726B} Birrer, S., Treu, T., Rusu, C.~E., et al.\ 2019, \mnras, 484, 4726. 
\bibitem[Birrer et al.(2016)]{2016JCAP...08..020B} Birrer, S., Amara, A., \& Refregier, A.\ 2016, \jcap, 2016, 020.
\bibitem[Cao et al.(2017)]{2017ApJ...835...92C} Cao, S., Li, X., Biesiada, M., et al.\ 2017, \apj, 835, 92. 
\bibitem[Liu et al.(2022)]{2022ApJ...927...28L} Liu, X.-H., Li, Z.-H., Qi, J.-Z., et al.\ 2022, \apj, 927, 28. 
\bibitem[Wei et al.(2022)]{2022ApJ...927L...1W} Wei, J.-J., Chen, Y., Cao, S., et al.\ 2022, \apjl, 927, L1. 
\bibitem[Liu \& Liao(2024)]{2024MNRAS.528.1354L} Liu, T. \& Liao, K.\ 2024, \mnras, 528, 1354. doi:10.1093/mnras/stae119
\bibitem[Eigenbrod et al.(2005)]{2005A&A...436...25E} Eigenbrod, A., Courbin, F., Vuissoz, C., et al.\ 2005, \aap, 436, 25. 
\bibitem[Treu et al.(2018)]{2018MNRAS.481.1041T} Treu, T., Agnello, A., Baumer, M.~A., et al.\ 2018, \mnras, 481, 1041. 
\bibitem[Birrer \& Treu(2021)]{2021A&A...649A..61B} Birrer, S. \& Treu, T.\ 2021, \aap, 649, A61. 
\bibitem[Shajib et al.(2023)]{2023A&A...673A...9S} Shajib, A.~J., Mozumdar, P., Chen, G.~C.-F., et al.\ 2023, \aap, 673, A9. 
\bibitem[Y{\i}ld{\i}r{\i}m et al.(2020)]{2020MNRAS.493.4783Y} Y{\i}ld{\i}r{\i}m, A., Suyu, S.~H., \& Halkola, A.\ 2020, \mnras, 493, 4783. 
\bibitem[Jee et al.(2016)]{2016JCAP...04..031J} Jee, I., Komatsu, E., Suyu, S.~H., et al.\ 2016, \jcap, 04, 031. 
\bibitem[Linder(2011)]{2011PhRvD..84l3529L} Linder, E.~V.\ 2011, \prd, 84, 123529. 
\bibitem[Suyu et al.(2017)]{2017MNRAS.468.2590S} Suyu, S.~H., Bonvin, V., Courbin, F., et al.\ 2017, \mnras, 468, 2590.
\bibitem[Shajib et al.(2020)]{2020MNRAS.494.6072S} Shajib, A.~J., Birrer, S., Treu, T., et al.\ 2020, \mnras, 494, 6072. 
\bibitem[Larkin et al.(2006)]{2006NewAR..50..362L} Larkin, J., Barczys, M., Krabbe, A., et al.\ 2006, \nar, 50, 362.
\bibitem[Birkmann et al.(2022)]{2022A&A...661A..83B} Birkmann, S.~M., Ferruit, P., Giardino, G., et al.\ 2022, \aap, 661, A83.
\bibitem[Wright et al.(2016)]{2016SPIE.9909E..05W} Wright, S.~A., Walth, G., Do, T., et al.\ 2016, Society of Photo-Optical Instrumentation Engineers (SPIE) Conference Series, 9909, 990905. 
\bibitem[Ivezi{\'c} et al.(2019)]{2019ApJ...873..111I} Ivezi{\'c}, {\v{Z}}., Kahn, S.~M., Tyson, J.~A., et al.\ 2019, \apj, 873, 111. 
\bibitem[Liao et al.(2015)]{2015ApJ...800...11L} Liao, K., Treu, T., Marshall, P., et al.\ 2015, \apj, 800, 11. 
\bibitem[Oguri \& Marshall(2010)]{2010MNRAS.405.2579O} Oguri, M. \& Marshall, P.~J.\ 2010, \mnras, 405, 2579. 
\bibitem[Schneider \& Sluse(2013)]{2013A&A...559A..37S} Schneider, P. \& Sluse, D.\ 2013, \aap, 559, A37. 
\bibitem[Schneider \& Sluse(2014)]{2014A&A...564A.103S} Schneider, P. \& Sluse, D.\ 2014, \aap, 564, A103. 
\bibitem[Abbott et al.(2017a)]{Abbott17} Abbott, B. P., et al. [LIGO Scientific Collaboration, the Virgo Collaboration], 2017a, \prl, 119, 161101
\bibitem[Abbott et al.(2017b)]{2017ApJ...848L..12A} Abbott, B.~P., Abbott, R., Abbott, T.~D., et al.\ 2017b, \apjl, 848, L12. 
\bibitem[Abbott et al.(2016)]{Abbott16} Abbott, B. P., et al. [LIGO Scientific and Virgo Collaborations], 2016, \prl, 116, 061102
\bibitem[Dalal et al.(2006)]{Dalal2006} Dalal, N., Holz, D. E., Hughes, S.A., Jain, B. 2006, PRD, 74, 063006
\bibitem[Taylor, et al.(2012)]{Taylor12} Taylor, S. R., et al. 2012, \prd, 85, 023535
\bibitem[Abernathy et al.(2011)]{Abernathy11} Abernathy, M., et al., ``Einstein gravitational wave Telescope: Conceptual Design Study", 2011 [available from European Gravitational Observatory, document number ET-0106A-10].
\bibitem[Taylor \& Gair(2012)]{Taylor12b} Taylor, S. R. \& Gair, J. R. 2012, \prd, 86, 023502
\bibitem[Califano et al.(2023a)]{2023MNRAS.518.3372C} Califano, M., de Martino, I., Vernieri, D., et al.\ 2023a, \mnras, 518, 3372. 
\bibitem[Califano et al.(2023b)]{2023PhRvD.107l3519C} Califano, M., de Martino, I., Vernieri, D., et al.\ 2023b, \prd, 107, 123519.
\bibitem[Cai \& Yang(2017)]{Cai17} Cai, R.-G. \& Yang, T. 2017, \prd, 95, 044024
\bibitem[Hirata et al.(2010)]{2010PhRvD..81l4046H} Hirata, C.~M., Holz, D.~E., \& Cutler, C.\ 2010, \prd, 81, 124046.
\bibitem[Tamanini et al.(2016)]{2016JCAP...04..002T} Tamanini, N., Caprini, C., Barausse, E., et al.\ 2016, \jcap, 2016, 002.

\bibitem[Speri et al.(2021)]{2021PhRvD.103h3526S} Speri, L., Tamanini, N., Caldwell, R.~R., et al.\ 2021, \prd, 103, 083526.
\bibitem[Hjorth et al.(2017)]{2017ApJ...848L..31H} Hjorth, J., Levan, A.~J., Tanvir, N.~R., et al.\ 2017, \apjl, 848, L31. 
\bibitem[Howlett \& Davis(2020)]{2020MNRAS.492.3803H} Howlett, C. \& Davis, T.~M.\ 2020, \mnras, 492, 3803. 
\bibitem[Mukherjee et al.(2021)]{2021A&A...646A..65M} Mukherjee, S., Lavaux, G., Bouchet, F.~R., et al.\ 2021, \aap, 646, A65. 
    2001, \mnras, 324, 797.

\bibitem[Kocsis et al.(2006)]{2006ApJ...637...27K} Kocsis, B., Frei, Z., Haiman, Z., et al.\ 2006, \apj, 637, 27.
\bibitem[Cen \& Ostriker(2000)]{2000ApJ...538...83C} Cen, R. \& Ostriker, J.~P.\ 2000, \apj, 538, 83. 
\bibitem[Schneider et al.(2001)]{2001MNRAS.324..797S} Schneider, R., Ferrari, V., Matarrese, S., et al.\ 2001, \mnras, 324, 797. 

\bibitem[Gao et al.(2022)]{2022MNRAS.516.1977G} Gao, R., Li, Z., \& Gao, H.\ 2022, \mnras, 516, 1977.

\bibitem[Capozziello et al.(2009)]{2009PhLB..671..193C} Capozziello, S., Elizalde, E., Nojiri, S., et al.\ 2009, Physics Letters B, 671, 193. 

\bibitem[More et al.(2017)]{2017MNRAS.465.2411M} More, A., Lee, C.-H., Oguri, M., et al.\ 2017, \mnras, 465, 2411.

\bibitem[Collett(2015)]{2015ApJ...811...20C} Collett, T.~E.\ 2015, \apj, 811, 20. 

\bibitem[Liao et al.(2022)]{2022ChPhL..39k9801L} Liao, K., Biesiada, M., \& Zhu, Z.-H.\ 2022, Chinese Physics Letters, 39, 119801. 
\bibitem[Liao et al.(2017)]{2017NatCo...8.1148L} Liao, K., Fan, X.-L., Ding, X., et al.\ 2017, Nature Communications, 8, 1148. 
\bibitem[Kelly et al.(2015)]{2015Sci...347.1123K} Kelly, P.~L., Rodney, S.~A., Treu, T., et al.\ 2015, Science, 347, 1123. 















\end{thebibliography}
\end{document}